\documentclass[showpacs]{revtex4}
\usepackage{graphicx}
\usepackage{bm}                 
\usepackage{amsmath}            
\usepackage{amsfonts}
\usepackage{amssymb}

\begin{document}
%
\title{Universal quantum computation using the discrete time quantum walk}
%
\author{Neil B. Lovett}
\email{pynbl@leeds.ac.uk}
\author{Sally Cooper}
\author{Matthew Everitt}
\author{Matthew Trevers}
\author{Viv Kendon}
\email{V.Kendon@leeds.ac.uk}
\affiliation{School of Physics and Astronomy, University of Leeds, Leeds, LS2 9JT,
 UK}

\begin{abstract}
A proof that continuous time quantum walks are universal for quantum computation, using unweighted graphs of low degree, has recently been presented by \citeauthor{childs09a} [PRL 102 180501 (2009)]. We present a version based instead on the discrete time quantum walk. We show the discrete time quantum walk is able to implement the same universal gate set and thus both discrete and continuous time quantum walks are computational primitives. Additionally we give a set of components on which the discrete time quantum walk provides perfect state transfer.

\end{abstract}
%
\maketitle              

\section{Introduction}
\label{sec:intro}
Quantum computers offer the promise of fundamentally faster processing based upon quantum mechanical properties. Although a physical device of a useful size is still to be built, many quantum algorithms have already been discovered. The most important of these are the algorithms introduced by \citeauthor{shor97a} and \citeauthor{grover96a}, which can factor integers and search an unsorted database respectively, significantly faster than the best known classical algorithms \cite{shor97a, grover96a}.

Quantum walks were initially introduced in both continuous \cite{farhi98a} and discrete \cite{aharonov93a} time, in direct analogy with their classical counterparts, and have since been studied extensively \cite{kempe03b}. In the same way that classical random walks are used in computer science for algorithm design, many quantum algorithms have been developed based upon quantum walks, with varying speed ups over the best known classical algorithms for the same problem, \cite{ambainis04b}. These solve the problems using two different approaches: hitting times and searching. In hitting time problems we start from a specific vertex and want to get to another as quickly as possible. These problems have yielded the largest speed up, including exponential speed ups over the classical case, \cite{childs03b, kempe03a, childs07a}. Searching for an entry in an unsorted database is a classically time-consuming problem taking on average a time of $O(N)$ to search a set of $N$ entries. Grover's algorithm, \cite{grover96a}, improves on this to $O(\sqrt{N})$ by using a technique known as amplitude amplification. The same speed up can be obtained using a quantum walk method on various structures \cite{shenvi03a, childs03a, childs04a}. 

In \cite{childs09a}, \citeauthor{childs09a} extends the original results of \citeauthor{feynman85a} \cite{feynman85a} to show a continuous time quantum walk, on an unweighted graph of bounded degree, is universal for quantum computation. \citeauthor{childs09a} gives an explicit construction that converts a standard gate model computation into a graph, on which a continuous time quantum walk executes an algorithm by traversing the graph. In this paper, we show the equivalent construction of a universal gate set using the discrete time quantum walk in place of the continuous. This confirms that both the continuous and the discrete time quantum walks can be regarded as computational primitives. The construction requires an exponentially large graph for the size of the input as we require $2^{n}$ wires for an $n$ qubit input. The quantum walk takes place on this $N$-vertex graph just as the continuous time walk does in the construction by Childs \cite{childs09a}. It is already known that a quantum walk on an $N$-vertex graph can be simulated efficiently by a universal quantum computer using $poly(\log N)$ gates, provided there is a simple rule to compute the neighbours of any vertex \cite{childs03b}. Thus, by performing the quantum walk on a quantum computer, the binary encoding brings the resources required back to the expected level.
 
Our construction for the universal gate set in discrete time is similar to \cite{childs09a} but has maximum degree, $d$, of eight at any vertex as opposed to three in the continuous case. The continuous time walk can easily be propagated in one direction with no reflection at the vertices. The discrete time walk is not so straightforward, it can only be propagated in one direction by using a specific coin corresponding to the $\sigma_{x}$ operation. Using this coin restricts the graph to vertices of degree two, providing no way to construct higher degree structures. Thus we must use a double-edged wire to accomplish directional propagation. This solution has its roots in the connection between the continuous and discrete time walks. \citeauthor{strauch06a} \cite{strauch06a} has shown that, as we take the continuous limit of the discrete time walk on the line, we actually get two copies of the continuous time walk propagating in opposite directions. \citeauthor{childs08a} \cite{childs08a} later showed a direct correspondence between the discrete and continuous time quantum walks on arbitrary graphs. In the same work, \citeauthor{childs08a} shows how a discrete time walk can be used, at its limit of small eigenvalues,  to approximate the continuous time walk. He uses this `lazy' quantum walk approach to allow the discrete time walk to propagate in the same way as the continuous. This same approach could be used in this case to allow the computation to be performed on the same structures defined in \cite{childs09a}. However, this would require the discrete time walk to approach the limit at which it is doing very little at each timestep. This would then increase the overhead required to allow completely deterministic computation.

We begin by describing the discrete time quantum walk briefly in section \ref{sec:discrete}, then move on to show structures on which the discrete time quantum walk will allow perfect state transfer in section \ref{sec:char}. These structures allow us to construct the elements we need to perform computation. Section \ref{sec:gate} shows the universal gate set we choose and how we implement these using the discrete time walk. Section \ref{sec:circuits} describes how we can link these gates and structures together to form any quantum circuit, and elaborates on how this is efficient, despite the size of the graph being exponential in the number of gates required. Finally, in section \ref{sec:conc} we discuss our findings and the differences with the continuous time construction of \cite{childs09a}.

\section{Discrete time Quantum Walk}
\label{sec:discrete}

Consider a classical random walk on a line in which a walker starts at a specific position and, depending on the outcome of a coin toss, moves either left or right. The outcome after many runs is a binomial distribution about the starting position with a spread (quantified by the standard deviation) of $\sqrt{t}$ where $t$ is the number of timesteps. A discrete time quantum walk is the direct analog of the classical walk with the walker replaced by a quantum particle carrying a two state quantum system for the coin. The coin toss is effected by a unitary operator. Although this is now deterministic, if we were to measure the coin we would get a random output as in the classical case. We start the quantum walker at the origin and act upon it with a unitary operator for the coin toss, followed by a conditional shift operation (to obtain the movement of the walker) at each timestep. We write the basis states of the walker as an ordered pair, $\mid x,c \rangle$, denoting the position of the walker, $x$, and the state of the coin, i.e., heads ($c = 1$) or tails ($c = 0$). The simplest unitary operator is the Hadamard operator, H, which acts on the state as
\begin{align}
H \mid x, 0 \rangle &= \frac{1}{\sqrt{2}}(\mid x, 0 \rangle + \mid x, 1 \rangle) \nonumber \\
H \mid x, 1 \rangle &= \frac{1}{\sqrt{2}}(\mid x, 0 \rangle - \mid x, 1 \rangle).
\label{hadamard}
\end{align}
The shift operation, S, acts thus
\begin{align}
S \mid x, 0 \rangle &= \mid x-1, 0 \rangle \nonumber \\
S \mid x, 1 \rangle &= \mid x+1, 1 \rangle.
\label{shift}
\end{align}
The first three timesteps starting at the origin are as follows
\begin{align}
(SH)^3\mid 0, 0 \rangle &=  (SH)^2 S \frac{1}{\sqrt{2}}(\mid 0, 0 \rangle \ + \mid 0, 1 \rangle) \nonumber \\
&= (SH)^2 \frac{1}{\sqrt{2}}(\mid -1, 0 \rangle \ + \mid 1, 1 \rangle) \nonumber \\
&= (SH) S \frac{1}{2}(\mid -1, 0 \rangle \ + \mid -1, 1 \rangle \ + \mid 1, 0 \rangle \ - \mid 1, 1 \rangle) \nonumber \\
&=  SH \frac{1}{2}(\mid -2, 0 \rangle \ + \mid 0, 1 \rangle \ + \mid 0, 0 \rangle \ - \mid 2, 1 \rangle) \nonumber \\
&= S \frac{1}{\sqrt{8}}(\mid -2, 0 \rangle \ + \mid -2, 1 \rangle \ + \mid 0, 0 \rangle \ - \mid 0, 1 \rangle \ + \mid 0, 0 \rangle \ + \mid 0, 1 \rangle \ - \mid 2, 0 \rangle \ + \mid 2, 1 \rangle) \nonumber \\
&= \frac{1}{\sqrt{8}}(\mid -3, 0 \rangle \ + \mid -1, 1 \rangle \ + 2 \mid -1, 0 \rangle \ - \mid 1, 0 \rangle \ + \mid 3, 1 \rangle).
\label{3steps}
\end{align}
As the walk progresses, quantum interference occurs whenever there is more
than one possible path of $t$ steps to the position.
This can be both constructive and destructive, as shown in eq.~(\ref{3steps}),
which causes some probabilities to be amplified or decreased at
each timestep. The walk on the line has been solved analytically \cite{ambainis01a, nayak00a} where it was first remarked that the quantum walk spreads quadratically faster than the classical one.

The choice of operator at each vertex can greatly affect the dynamics of the walk and its propagation across the structure. A bias can be introduced \cite{bach02a}, for $d=2$, this is done using a generalisation of the Hadamard operator,

\begin{equation}
H_{bias} = \left ( \begin{matrix} \sqrt{\delta} & \sqrt{1-\delta} \\ \sqrt{1-\delta} & -\sqrt{\delta} \end{matrix} \right ),
\end{equation}
where $\delta$ is the bias in the coin. Setting this to $\delta=\frac{1}{2}$ returns the standard Hadamard operator, eq.~(\ref{hadamard}). Similarly, the choice of the walker's initial state is also important, unlike in the classical random walk. A good review of these effects can be found in \cite{kendon03a}. 

\begin{figure}[!tb]
\begin{minipage}{\columnwidth}
    \centering
	\includegraphics[scale=0.4]{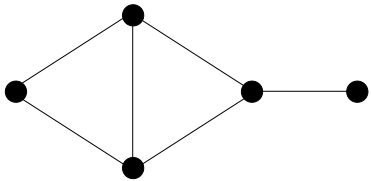}
	\caption{A general graph, G, with 5 vertices and 6 edges. It is undirected and has vertices of varying degree.}
    \label{graph}
\end{minipage}
\end{figure}

For universal computation we need a quantum walk on a more complex graph. In graph theory, a general graph, G, is an ordered pair consisting of a set of vertices, V, and a set of edges, E, which link the vertices. The number of edges incident on a vertex is the degree of the vertex, fig.~\ref{graph} shows a small general graph which has vertices of varying degree.
It is also undirected, meaning that the edges allow movement in both directions. A different operator (coin) is needed at vertices with $d>2$ in order to act on the entire state space \cite{mackay02a,kendon06a}. 

The Grover coin can be extended to any degree at a vertex,

\begin{equation}
G^{(d)} = \left[ \begin{matrix}  \frac{2}{d} & \dots &  \frac{2}{d} \\  \vdots & \ddots & \vdots \\  \frac{2}{d} & \dots &  \frac{2}{d} \end{matrix} \right ] - I_{d},
\label{grover}
\end{equation}
where $d$ is the degree of the vertex and $I_{d}$ is the identity matrix of the same dimension. 

Coins of degree four will be needed at most of the vertices in our computational graphs. The Grover coin in four dimensions, eq.~(\ref{grover}), reduces to
\begin{equation}
G^{(4)} = \frac{1}{2}\left[ \begin{matrix}  -1 & 1 & 1 & 1 \\ 1 & -1 & 1 & 1 \\ 1 & 1 & -1 & 1 \\ 1 & 1 & 1 & -1 \end{matrix} \right ].
\label{grover4}
\end{equation}
\begin{figure}[!tb]
\begin{minipage}{\columnwidth}
    \centering
	\includegraphics[scale=0.9]{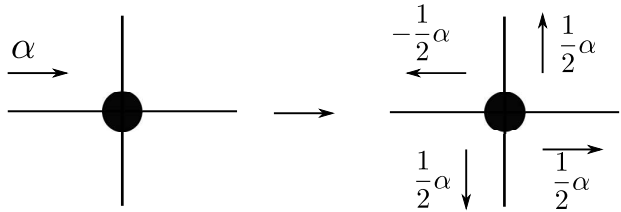}
	\caption{An example of a portion of the walker reflecting back upon itself. This is a single degree 4 vertex with the Grover coin, eq.~(\ref{grover}), operating on the incoming amplitude $\alpha$.}
    \label{reflection}
\end{minipage}
\end{figure}
\\*Using these higher dimensional coins can cause the walker to be reflected back upon itself with some probability. Figure \ref{reflection} shows this reflection for a vertex of degree $d=4$. In quantum walk search and other quantum walk algorithms this can be useful to provide interference. However, here we need to ensure the walker moves in one direction only, from left to right, so it most resembles the circuit model of quantum computation. We show how we accomplish this forward only propagation in section \ref{sec:gate}. 
 
\section{Perfect State Transfer}
\label{sec:char}

As a preliminary to our quantum computation scheme, we discuss structures on which perfect state transfer can be achieved using the discrete time quantum walk. Perfect state transfer has been investigated in the context of spin chains by \citeauthor{landahl04a} \cite{landahl04a, landahl05a}. The propagation of the state through spin systems follows the same dynamics as a continuous time quantum walk. Perfect state transfer can occur on chains of length 2 or 3, hypercubes of any size, and chains with different coupling strengths engineered to optimise state transfer.

\begin{figure}[!tb]
\begin{minipage}{\columnwidth}
    \centering
	\includegraphics[scale=1]{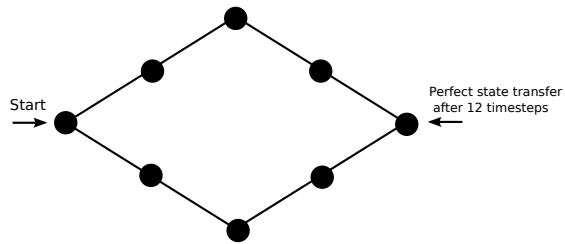}
	\caption{Cycle of 8 vertices which gives perfect state transfer from the initial vertex to the opposite vertex after half of the period, 12 timesteps. The entire state returns to the initial vertex in a full period, 24 timesteps.}
    \label{transfer}
\end{minipage}
\end{figure}

The closely related properties of instantaneous mixing and periodic cycles have been studied in detail for quantum walks. For the continuous time quantum walk, instantaneous mixing has been investigated by \citeauthor{tamon03a} \cite{tamon03a,tamon06a,tamon07a,tamon08a}. They showed \cite{tamon03a} this is achieved on cycles of 2, 3 and 4 vertices only. For the discrete time walk, slightly larger cycles show exact periodic behaviour. \citeauthor{travaglione02a}  \cite{travaglione02a} showed that a cycle of 4 vertices has a periodicity of 8 timesteps after which the entire state returns to the starting position. \citeauthor{kendon03a}   \cite{kendon03a} showed more periodic cycles exist, cycles of 2, 3, 4, 5, 6, 8 and 10 were shown numerically to be periodic by varying both the bias and phase in the coin. Perfect state transfer occurs at half the periodic cycle for even cycles, where we obtain the entire state at the opposite point of the cycle as shown in fig.~\ref{transfer}. 

For our case of using the walk for computation, we require the walk to travel perfectly in a single direction. On the structures mentioned, the quantum walk travels around the cycle in both directions and interferes to produce perfect state transfer. Using a completely biased coin,

\begin{equation}
H_{max. bias} = \left ( \begin{matrix} 0 & 1 \\ 1 & 0 \end{matrix} \right ),
\end{equation}
we can make the state transfer perfectly around the cycle in a single direction. However, if we then try to attach another structure to the cycle, this periodicity is broken in both cases. The Grover coin, eq.~(\ref{grover}), can be used to overcome part of this problem at vertices with an equal number of input edges as output edges. For any vertex of even degree it will transfer the entire state from all the input edges to all the output edges provided the inputs are all equal in both amplitude and phase. These results led us to the designs that work for universal computation.

\section{Universal Gate Set}
\label{sec:gate}

We now show how we construct a universal gate set with the discrete time quantum walk. Although the gate set we implement is the same as in \cite{childs09a}, the structures used to propagate the discrete time walk are different. The gate set used is the standard universal set comprising the controlled-not (C-NOT) gate,

\begin{equation}
C_{NOT} = \left [ \begin{matrix} 1 & 0 & 0 & 0 \\ 0 & 1 & 0 & 0 \\  0 & 0 & 0 & 1 \\ 0 & 0 & 1 & 0 \end{matrix} \right ],
\label{cnot}
\end{equation}
the single qubit Hadamard,

\begin{equation}
H = \frac{1}{\sqrt{2}} \left [ \begin{matrix} 1 & 1 \\ 1 & -1 \end{matrix} \right ],
\label{hadamardmatrix}
\end{equation}
and phase shift gates (we implement the specific phase shift known as the $\frac{\pi}{8}$ gate),

\begin{equation}
P(\frac{\pi}{8}) = \left [ \begin{matrix} 1 & 0 \\ 0 & e^{i\frac{\pi}{4}} \end{matrix} \right ].
\label{phase}
\end{equation}
These gates create a universal set that can implement any quantum computation \cite{barenco95a}. 

\begin{figure}[!tb]
\begin{minipage}{\columnwidth}
    \centering
	\includegraphics[scale=0.5]{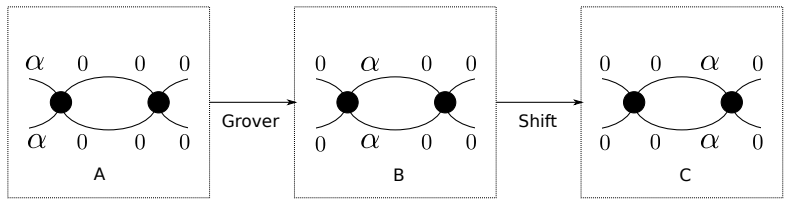}
	\caption{Grover and shift operation acting on a vertex of degree $d=4$. Section A shows the initial state, B shows the state after the Grover coin is applied and finally C is after the shift operation.}
    \label{prop}
\end{minipage}
\end{figure}

\begin{figure}[!tb]
\begin{minipage}{\columnwidth}
    \centering
	\includegraphics[scale=0.4]{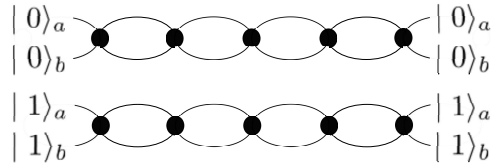}
	\caption{Basic wire used to propagate the quantum walk from left to right only. At a vertex of degree $d=4$ the Grover diffusion coin is used. The initial state is split across the pair of edges in the wire, eq.~(\ref{initialstate}).}
    \label{wire}
\end{minipage}
\end{figure}

In order to represent quantum states, \citeauthor{childs09a} defines his computational basis states as quantum wires. The other gates required for universality are then attached to wires and used to connect them together. The computation is represented as a quantum walk on these wires and structures, where the computation flows from input to output (left to right in our diagrams). Note that this encoding is not meant to be implemented directly. The wires represent computational basis states rather than qubits, thus the model does not represent a physical architecture. Instead, the underlying graph structure created would be used to help `program' a quantum computer. We first show how to construct a simple wire along which the quantum walk will propagate naturally in one direction. We use two edges per wire to ensure no reflection occurs at a vertex. We distribute the walker across the two edges which then recombine at the next vertex. As the split is equal, the Grover coin in effect moves both halves to the output edges of the vertex. Figure \ref{prop} shows this operation and the `shift' to the next vertex in explicit steps. The Grover diffusion coin, eq.~(\ref{grover}), is used at each vertex of degree $d=4$. The initial and final vertices are in effect degree four if we include other edges attached to either end. Figure \ref{wire} shows the basic wire we use. The computation would start with the amplitude at the initial vertex spread equally across the pair of edges in a wire. For example, the state $\alpha \mid 0 \rangle + \beta \mid 1 \rangle$, where $\mid \alpha \mid^{2} + \mid\beta\mid^{2} = 1$, would be split thus,

\begin{equation}
\mid \psi_{initial} \rangle = \frac{1}{\sqrt{2}} \left [ \alpha \mid 0 \rangle_{a} + \alpha \mid 0 \rangle_{b} + \beta \mid 1 \rangle_{a} + \beta \mid 1 \rangle_{b}  \right ],
\label{initialstate}
\end{equation}
where the subscript $a$ refers to the top line of the wire and subscript $b$ is the bottom line. The walk propagates left to right on the wire deterministically, in this case reaching the incoming edges of the final vertex in four timesteps. These wires form the basic connections in the computation.

\begin{figure}[!tb]
\begin{minipage}{\columnwidth}
    \centering
	\includegraphics[scale=0.4]{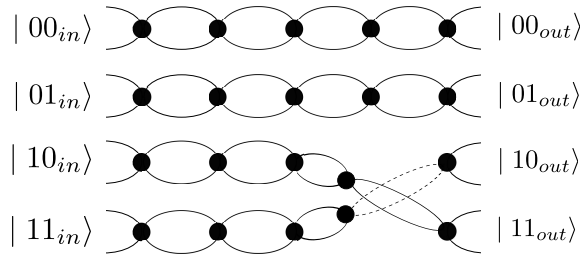}
	\caption{Structure used to implement a C-NOT gate. In this case the first qubit is the control and the second is the target. The target qubit's wires are interchanged and the control qubit is left unaltered. The dotted lines represent wires passing underneath the solid lines - there is no interaction between these wires.}
    \label{cnotwires}
\end{minipage}
\end{figure}

The simplest gate to construct is the C-NOT. It is trivial to implement by just exchanging the wires of the second qubit. The C-NOT gate is shown in fig.~\ref{cnotwires} and shows how the second qubit is flipped but the first qubit is untouched.

The phase gate, eq.~(\ref{phase}), requires the addition of a relative phase to one wire or computational basis state in relation to the other. To accomplish this, but still have only one coin operator for each vertex of the same degree, we modify the basic wire and add a phase factor, $e^{i\phi}$, to it,

\begin{equation}
G_{\phi}^{(4)} = e^{i\phi} G^{(4)}.
\label{phasecoin}
\end{equation}
Thus, as the walk propagates along a basic wire, it now picks up a phase of $e^{i\phi}$ each time it passes through a vertex of degree $d=4$. For the wires shown in fig.~\ref{wire} the walker would pick up a phase of $e^{5i\phi}$ as it reaches the final vertex.
\begin{figure}[!tb]
\begin{minipage}{\columnwidth}
    \centering
	\includegraphics[scale=0.4]{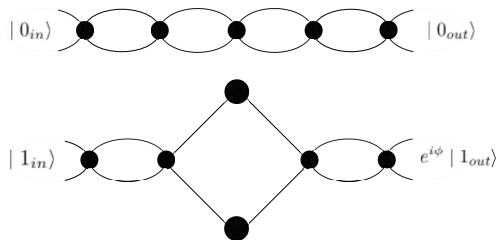}
	\caption{Phase gate structure. The $d=2$ Grover coin, eq.~(\ref{grover2}), is used at the vertices of degree $d=2$. The $\mid1\rangle$ wire will pick up a phase of $e^{i\phi}$ relative to the $\mid0\rangle$ wire. In our construction we actually obtain the operation corresponding to a phase of $e^{i\frac{\pi}{4}}$ as we set $\phi=-\pi/4$.}
    \label{phasewire}
\end{minipage}
\end{figure}
The phase added here is arbitrary and can be set to any value so long as it is set to the same value for all vertices of degree $d=4$. As we are looking to implement a $\pi/8$ gate we set it as follows: $\phi = -\pi/4.$ In order to add a relative phase of $\pi/4$ between the $\mid 0 \rangle$ and $\mid 1 \rangle$ wires we insert the structure in fig.~\ref{phasewire} into the graph at the required point. In this structure there are also vertices of degree $d=2$, at which we use the Grover coin at its limit of degree $d=2$,
\begin{equation}
G^{(2)} = \left [ \begin{matrix} 0 & 1 \\ 1 & 0 \end{matrix} \right ].
\label{grover2}
\end{equation}
We add no phase to eq.~(\ref{grover2}) so as the walker passes through these vertices no phase is picked up. In fig.~\ref{phasewire}, the walker propagates along the $\mid 1 \rangle$ wire and picks up a phase of $e^{-i\pi}$ in four timesteps as it only passes through four vertices of degree four. However, the $\mid 0 \rangle$ wire picks up a phase of $e^{-5i\frac{\pi}{4}}$ in the same number of timesteps as all its vertices are of degree four. Relative to the $\mid 0 \rangle$ wire, the $\mid 1 \rangle$ wire will pick up a phase of $e^{i\frac{\pi}{4}}$. Therefore, using the structure described here we obtain the operation in eq.~(\ref{phase}). 

\begin{figure}[!tb]
\begin{minipage}{\columnwidth}
    \centering
	\includegraphics[scale=0.25]{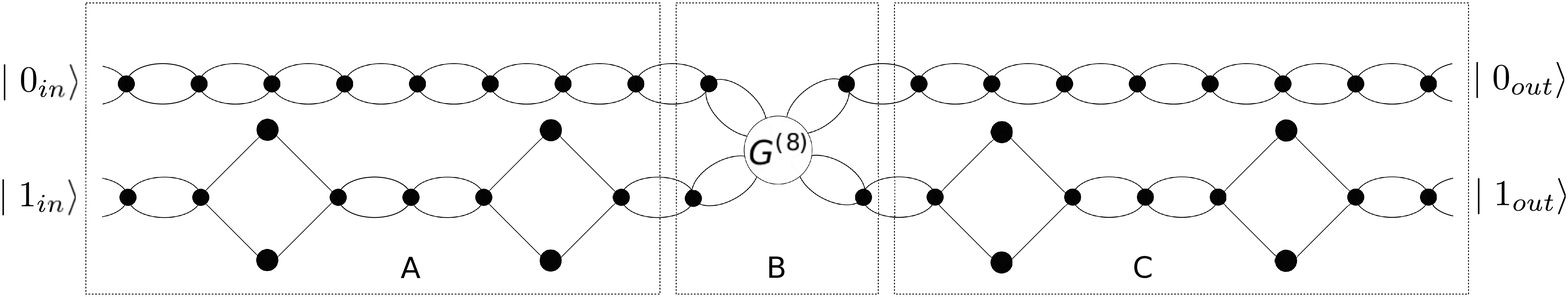}
	\caption{Hadamard gate structure. Sections A and C add a relative phase of $i$ to the $\mid1\rangle$ wire. The structure adds a global phase of $3\pi/4$ to the wires.}
    \label{hadamardwires}
\end{minipage}
\end{figure}

The last gate in the universal set is the Hadamard gate. This requires an interaction between the two computational basis states. The structure we use to perform this operation is shown in fig.~\ref{hadamardwires}. 
This looks complex in relation to the other gates we have shown so we break it up to explain it more clearly. Sections A and C of the structure are each two phase gates giving a relative phase of $i$ to the $\mid 1 \rangle$ wire before and after the main section of the gate (B). Section B of the structure combines the two inputs from the $\mid0\rangle$ and $\mid1\rangle$ wires and then splits this across the outputs equally. The structure here is similar to the basis changing gate in \cite{childs09a}. In order to obtain the desired operation on this structure we have designed a coin for vertices of degree $d=8$:
\begin{equation}
G^{(8)} = \frac{1}{2} \left [ \begin{matrix} 0 & 0 & 0 & 0 & 1 & i & i & -1 \\ 
0 & 0 & 0 & 0 & i & 1 & -1 & i \\
0 & 0 & 0 & 0 & i & -1 & 1 & i \\
0 & 0 & 0 & 0 & -1 & i & i & 1 \\
i & -1 & 1 & i & 0 & 0 & 0 & 0 \\
 -1 & i & i & 1 & 0 & 0 & 0 & 0 \\
1 & i & i & -1 & 0 & 0 & 0 & 0 \\
i & 1 & -1 & i & 0 & 0 & 0 & 0 \\
 \end{matrix} \right ].
\label{hadamardcoin}
\end{equation}
This operator combines the complex Hadamard operator,
\begin{equation}
H_{i} = \frac{1}{\sqrt{2}} \left [ \begin{matrix} 1 & i \\ i & 1 \end{matrix} \right ],
\label{complexhadamardmatrix}
\end{equation}
and the $\sigma_{x}$,
\begin{equation}
\sigma_{x} = \left [ \begin{matrix} 0 & 1 \\ 1 & 0 \end{matrix} \right ],
\label{paulizmatrix}
\end{equation}
in a tensor product form,
\begin{equation}
G^{(8)} = ( H_{i} \otimes H_{i} ) \otimes \sigma_{x},
\label{tensorg8}
\end{equation}
with the top two and bottom two rows of the $H_{i} \otimes H_{i}$ matrix rearranged. This rearrangement ensures the outputs come out in the same order as the input states. This gate adds a global phase of $3\pi/4$. The phase gates at the start and end of the central section give us the relative phase of $-1$ required for the Hadamard operation. We note here that the choice of where to place the phases in this construction is arbitrary. The same result can be achieved by using the degree four Grover coin, eq.~(\ref{grover4}), with no phase at vertices of degree four and the degree two Grover coin, eq.~(\ref{grover2}), with a phase of $\pi/4$ at vertices of degree two. However, by placing the phases on the Grover coin in the fashion we described previously, eqs.~(\ref{phasecoin}, \ref{grover2}), means that the global phase added by the Hadamard gate, eq.~(\ref{hadamardcoin}), corresponds to the phase added by a wire of the same length. 

\section{Constructing Quantum Circuits}
\label{sec:circuits}

\begin{figure}[!tb]
\begin{minipage}{\columnwidth}
    \centering
	\includegraphics[scale=0.6]{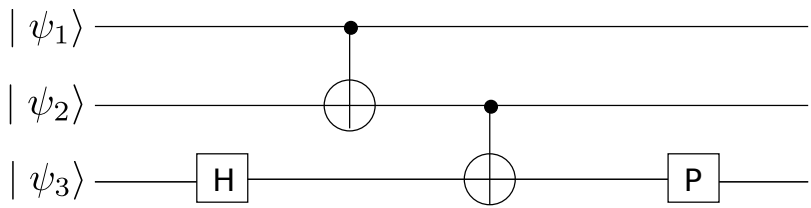}
	\caption{Quantum circuit on three qubits. A Hadamard operation is performed on qubit three followed by two C-NOT gates. Finally a phase gate is applied on qubit three. The underlying graph of this structure is shown in fig.~\ref{circuitgraph}}
    \label{qcircuit}
\end{minipage}
\end{figure}

\begin{figure}[!tb]
\begin{minipage}{\columnwidth}
    \centering
	\includegraphics[scale=0.5]{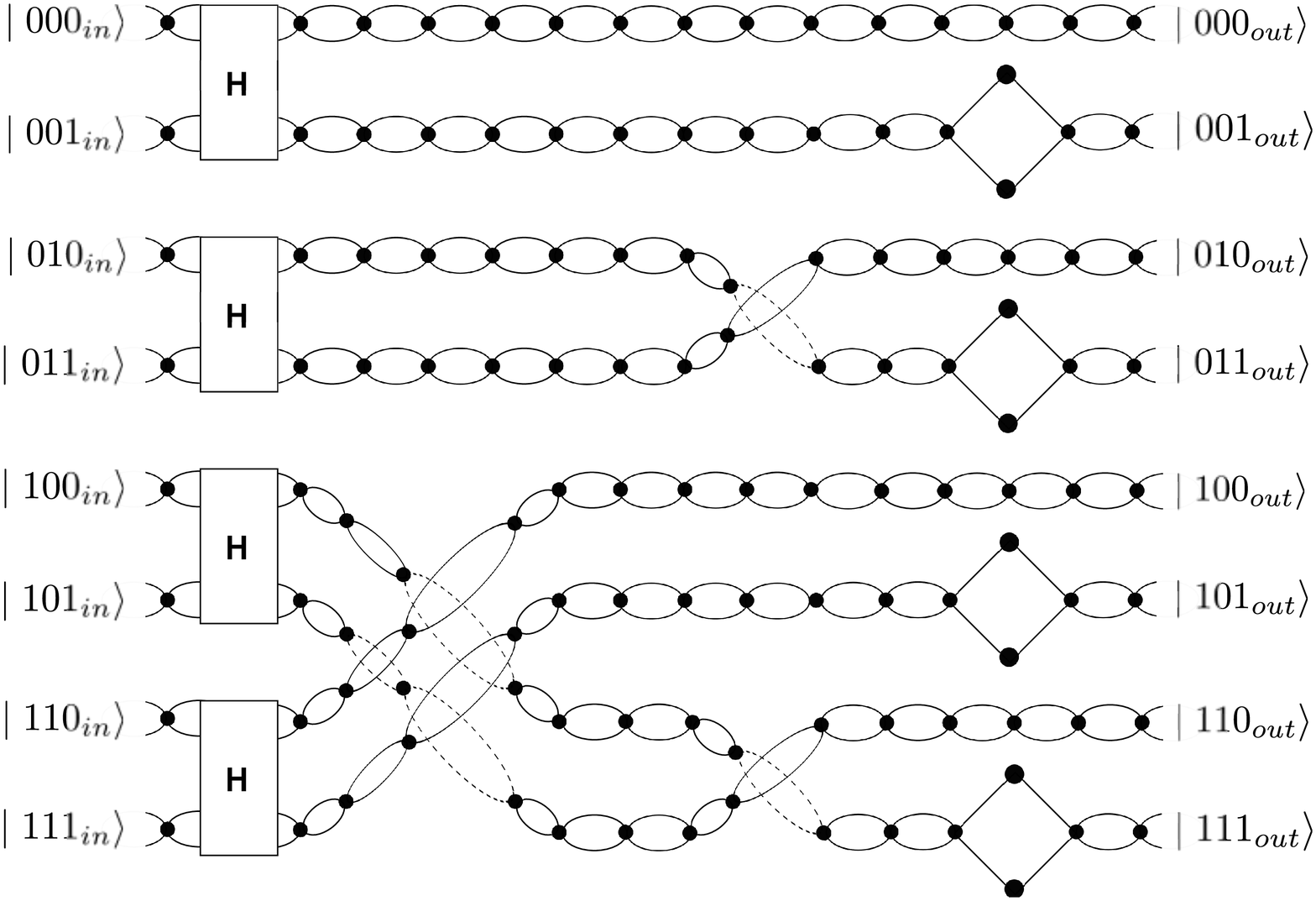}
	\caption{Graph to represent the quantum circuit in fig.~\ref{qcircuit}. The Hadamard structure, H, is the same as in fig.~\ref{hadamardwires}. The dotted lines represent wires passing underneath the solid lines - there is no interaction between these wires.}
    \label{circuitgraph}
\end{minipage}
\end{figure}

Thus far each gate we have described only acts on one or two qubits. However, non-trivial quantum computers involve many qubits. We now describe how to link these wires and structures together to form larger circuits. Figure~\ref{circuitgraph} shows the underlying graph structure of the circuit in fig.~\ref{qcircuit}. The graph structure is obtained by connecting together wires and structures so that the walk flows from left to right. For this reason, we designed our wire and structures with both input and output vertices of degree four, thus making it simple to link them together. The initial state of the computation is set on all or a subset of the vertices on the left hand side of the graph, with the amplitude at each vertex split across the incoming edges. This initial state can be thought of as the first column of vertices in the graph structure in superposition, with each subsequent column of vertices representing a further timestep. For example, in fig.~\ref{circuitgraph} this column of vertices is the set prior to the Hadamard structures. The walker is propagated across the graph structure, from left to right deterministically, for the required number of timesteps. We therefore do not require the addition of momentum filters or separators as in the continuous time case. Our structures all propagate the walker at the same speed, meaning output from the wires will be synchronised throughout the computation. Finally, the walker picks up a global phase of $-\pi/4$ per vertex that is not part of a gate that changes the phase, so all the wires also stay synchronized in phase. Thus, we know with certainty that, after the required number of timesteps, the walker will have a distribution over just the output vertices on the right hand side of the graph. Once the computation has been completed, we measure the output vertices. We will find the walker at just one of these vertices, representing the output of the computation.

The graph structure in fig.~\ref{circuitgraph} is clearly larger in size than its equivalent representation in the circuit model, fig.~\ref{qcircuit}. In fact, for a general $n$-qubit computation the equivalent graph will have $2^{n}$ wires, one for each combination of computational basis states. Similarly, we require more gate structures than in the circuit model. Single qubit structures are repeated $2^{n-1}$ times and for the C-NOT gate we need $2^{n-2}$ structures. As an example, we can see the phase gate acting on qubit 3 in fig.~\ref{qcircuit} is repeated 4 times in the underlying graph structure of fig.~\ref{circuitgraph}, one for each combination of wires involving qubit 3. Although this seems as though we would lose any form of quantum speed up due to the exponential number of gates required in the underlying graph, this is not the case. Consider simulating a classical random walk on a classical computer, the $N$-vertex graph is represented in $\log_{2} N$ bits of memory with each vertex having a unique binary number as a label. In a similar fashion, if we simulate a quantum walk on a quantum computer, the $N$-vertex graph can be represented by $\log_{2} N$ qubits. Therefore if we encode our graph using qubits, we can describe the $2^{n}$ wires in just $n$ qubits. By manipulation of a single qubit we can affect all combinations of wires associated with that qubit. As the state moves across the graph, the adjacent vertices must be established. In complex graphs the description of the graph and its connections is often exponential in size and an oracle must be used to store it \cite{childs03b}. The graphs produced here are of bounded degree and have a regularity stemming from the repetition of gate structures on combinations of wires involving a specific qubit. Due to the labelling of the wires, we know where to place each structure based on one bit in the label, thus we can efficiently describe the graph. For example, consider the second C-NOT gate in fig.~\ref{qcircuit}, which operates on qubit 3 with qubit 2 as control. We can see from fig.~\ref{circuitgraph} that it is easy to identify where the C-NOT structures should be placed. The labelling of the wires shows that the middle bit determines which combinations of wires relate to the 2nd qubit having a value of 1. Similarly, we can also identify which wire it should link to by the last bit in the labelling scheme, it needs to be flipped relative to the original wire, i.e. $\mid 010 \rangle$ links to $\mid 011 \rangle$. 

\section{Discussion}
\label{sec:conc}

In this paper we have described an alternative to the construction in \cite{childs09a} using the discrete time quantum walk. This shows the discrete time quantum walk is universal, therefore any quantum algorithm can be reformulated as a discrete time quantum walk algorithm. It also confirms that the discrete and continuous time walks are both computational primitives and thus computationally equivalent. This equivalence is dependent on the number of steps in both cases to be of the same order. Our gate constructs require twice the number of edges compared to the continuous time case but the same number of wires. Our phase gate requires an additional timestep in relation to the continuous time phase gate construct. The number of timesteps required for a computation is also the same as the continuous time case but with a small overhead depending on the number of phase gates required.

Another difference in the two constructions is the degree of the graphs produced. In the continuous time case the maximum degree of any vertex in the graph is three. In the discrete time case we use vertices of higher degree to ensure directional propagation. In most of the structures this is a doubling of the degree at a vertex, as shown in the case of the basic wire and the phase gate structure. The Hadamard structure we propose here however, does not follow this doubling. It would seem reasonable, from the equivalent degree three structure in the work by \citeauthor{childs09a}, that it may be possible to decompose our Hadamard structure into one with degree six vertices. The doubling of degree at vertices would then correspond directly to the continuous case.

\textit{Acknowledgments:}
We thank Andrew Childs for helpful comments on a draft of
the manuscript.
NL is funded by the UK Engineering and Physical Sciences Research Council.
VK is funded by a Royal Society University Research Fellowship.
SC and MT were funded by Nuffield Foundation Science Undergraduate Research Bursaries.
ME was funded by the University of Leeds.

%
%


\begin{thebibliography}{99}

\bibitem[Shor(1997)]{shor97a}
P.~W.~Shor, SIAM.~J.~Comput., \textbf{26}, 1484, (1997).

\bibitem[Grover(1996)]{grover96a}
L.~K.~Grover, in \textit{Proceedings~of~the~28th~Annual~ACM~STOC, 1996} (ACM, New York, 1996), pp.~212.

\bibitem[Farhi and Gutmann(1998)]{farhi98a}
E.~Farhi and S.~Gutmann, Phys.~Rev.~A, \textbf{58}, 915, (1998).

\bibitem[Arahonov (1993)]{aharonov93a}
Y.~Aharonov, L.~Davidovich and N.~Zagury
\newblock \emph{Phys.~Rev.~A}, \textbf{48}, 1687, 1993.

\bibitem[Kempe(2003)]{kempe03b}
J.~Kempe, Contempory~Physics, \textbf{44}, 307, (2003).

\bibitem[Ambainis(2003)]{ambainis04b}
A.~Ambainis, Intl.~J.~ Quant.~Info., \textbf{1} (4), 507, (2003).

\bibitem[Childs(2003)]{childs03b}
A.~M.~Childs, R.~Cleve, E.~Deotto, E.~Farhi, S.~Gutmann and D.~Spielman, \textit{Proc.~35th~ACM~Symposium~on~Theory~of~Computing}, pp.~59-68, (2003).

\bibitem[Kempe(2003)]{kempe03a}
J.~Kempe, \textit{Proc.~of~RANDOM'03~Lecture~Notes~in~Computer~Science}, 2764, pp.~354-369, (2003).

\bibitem[Childs(2007)]{childs07a}
A.~M.~Childs, L.~J.~Schulman and U.~V.~Vazirani, \textit{Proc.~48th~IEEE~Symposium~on~Foundations~of~Computer~Science}, pp.~395-404, (2007).

\bibitem[Shenvi~et~al.(2003)]{shenvi03a}
N.~Shenvi, J.~Kempe and K.~Whaley, Phys.~Rev.~A, \textbf{67}, 052307, (2003).

\bibitem[Childs(2003)]{childs03a}
A.~M.~Childs and J.~Goldstone, Phys.~Rev.~A, \textbf{70}, 022314, (2004).

\bibitem[Childs(2004)]{childs04a}
A.~M.~Childs and J.~Goldstone, Phys.~Rev.~A, \textbf{70}, 042312, (2004).

\bibitem[Childs(2009)]{childs09a}
A.~M.~Childs, Phys.~Rev.~Lett., \textbf{102}, 180501, (2009).

\bibitem[Feynman(1985)]{feynman85a}
R.~P.~Feynman, Optics~News, \textbf{11}, 11, (1985).

\bibitem[Strauch(2006)]{strauch06a}
F.~W.~Strauch, Phys.~Rev.~A, \textbf{74}, 030301, (2006).

\bibitem[Childs(2008)]{childs08a}
A.~M.~Childs, Communications.~in.~Math.~Phys.,\textbf{294}, 581, (2010).

\bibitem[Ambainis(2001)]{ambainis01a}
A.~Ambainis, E.~Bach, A.~Nayak, A.~Vishwanath and J.~Watrous, \textit{Proc.~33rd Annual ACM Symposium on Theory of Computing}, pp.~60--69, (2001).

\bibitem[Nayak(2000)]{nayak00a}
A.~Nayak and A.~Vishwanath, quant-ph/0010117, (2000).

\bibitem[Bach et al.(2002)]{bach02a}
E.~Bach, S.~Coppersmith, M.~Paz-Goldschen, R.~Joynt and J.~Watrous, J.~Comput.~Syst.~Sci., \textbf{69} (4), 562, (2004).

\bibitem[Tregenna et al.(2003)]{kendon03a}
B.~Tregenna, W.~Flanagan, R.~Maile and V.~Kendon, New.~J.~Phys., \textbf{5}, 83, (2003).

\bibitem[Mackay et al.(2002)]{mackay02a}
T.~D.~Mackay, S.~D.~Bartlett, L.~T.~Stephenson and B.~C.~Sanders, J.~Phys.~A:~Math.~Gen., \textbf{35}, 2745, (2002).

\bibitem[Kendon(2006)]{kendon06a}
V.~Kendon, Intl.~J.~ Quant.~Info., \textbf{4} (5), 791, (2006).

\bibitem[Travaglione and Milburn(2002)]{travaglione02a}
B.~C.~Travaglione and G.~J.~Milburn, Phys.~Rev.~A, \textbf{65}, 032310, (2002).

\bibitem[Tamon et al.(2003)]{tamon03a}
A.~Ahmadi, R.~Belk, C.~Tamon and C.~Wendler, Quant.~Info.~and~Comp., \textbf{3}, 611, (2003).

\bibitem[Fedichkin et al.(2006)]{tamon06a}
L.~Fedichkin, D.~Solenov and C.~Tamon, Quant.~Info.~and~Comp., \textbf{6} (3), 263, (2006).

\bibitem[Carlson et al.(2007)]{tamon07a}
W.~Carlson, A.~Ford, E.~Harris, J.~Rosen, C.~Tamon and K.~Wrobel, Quant.~Info.~and~Comp., \textbf{7} (8), 738, (2007).

\bibitem[Best et al.(2008)]{tamon08a}
A.~Best, M.~Kliegl, S.~Mead-Gluchacki and C.~Tamon, Intl.~J.~ Quant.~Info., \textbf{6} (6), 1135, (2008).

\bibitem[Landahl et al.(2004)]{landahl04a}
M.~Christandl, N.~Datta, A.~Ekert and A.~J.~Landahl, Phys.~Rev.~Lett., \textbf{92}, 187902, (2004).

\bibitem[Christandl et al.(2005)]{landahl05a}
M.~Christandl, N.~Datta, T.~C.~Dorlas, A.~Ekert, A.~Kay and A.~J.~Landahl, Phys.~Rev.~A, \textbf{71}, 032312, (2005).

\bibitem[Barenco et al.(1995)]{barenco95a}
A.~Barenco, C.~H.~Bennett, R.~Cleve, D.~P.~DiVincenzo, N.~Margolus, P.~Shor, T.~Sleator, J.~Smolin and H.~Weinfurter, Phys.~Rev.~A, \textbf{52}, 3457, (1995).

\end{thebibliography}
\end{document}